\documentclass[runningheads]{llncs}

\usepackage{makecell}
\usepackage{graphicx}
\usepackage{hyperref}
\usepackage{float}
\usepackage{siunitx}
\usepackage{amssymb}
\usepackage{amsmath}
\usepackage{multirow}
\usepackage{bbding}
\usepackage[marginal]{footmisc}

\begin{document}
\title{CTooth+: A  Large-scale Dental Cone Beam Computed Tomography Dataset and Benchmark for Tooth Volume Segmentation}

\titlerunning{CTooth+}
%
%

\authorrunning{Weiwei Cui et al.}

\author{Weiwei Cui\inst{2}, Yaqi Wang \inst{1}\inst{(}\Envelope\inst{)},  Yilong Li\inst{2}, Dan Song \inst{4}, Xingyong Zuo \inst{4}, Jiaojiao Wang \inst{1}, Yifan Zhang\inst{3},  Huiyu Zhou \inst{5}, Bung san Chong\inst{2}, Liaoyuan Zeng  \inst{4}, Qianni Zhang \inst{2}\inst{(}\Envelope\inst{)}}

\authorrunning{Weiwei et al.}
%
\institute{Communication University of Zhejiang\\
\and
Queen Mary University of London \\
\and 
West China Hospital of Stomatology, Sichuan University \\
\and
University of Electronic Science and Technology of China\\
\and
University of Leicester
} 
%

%
\maketitle              
\begin{abstract}
Accurate tooth volume segmentation is a prerequisite for computer-aided dental analysis. Deep learning-based tooth segmentation methods have achieved satisfying performances but require a large quantity of tooth data with ground truth. The dental  data  publicly available is limited meaning the existing methods can not  be reproduced, evaluated and applied in clinical practice.  In this paper, we establish a 3D dental CBCT dataset CTooth+, with 22 fully annotated volumes and 146 unlabeled volumes. We further evaluate several state-of-the-art tooth volume segmentation strategies based on fully-supervised learning, semi-supervised learning and active learning, and define the performance principles. This work provides a new benchmark for the tooth volume segmentation task, and the experiment can serve as the baseline for future AI-based dental imaging research and clinical application development. The codebase and dataset are released \href{https://github.com/liangjiubujiu/CTooth}{here}.

\keywords{3D segmentation   \and dental dataset \and fully supervised learning \and semi-supervised learning \and active learning.}
\end{abstract}
\section{Introduction}
Accurately segmented tooth volumes  provide  valuable 3D information for the clinical diagnosis such as  root shape, curvature, tooth  size, the spatial relationship of multiple teeth. However, manually delineating all tooth regions is labour-consuming, error-prone and expensive.

Some learning-based methods have been proposed to automatically segment tooth regions and achieve approving results. Several shallow learning-based methods try to quickly segment teeth from X-ray or CBCT images such as region-based \cite{lurie2012recursive}, threshold-based \cite{ajaz2013dental}, and cluster-based \cite{alsmadi2018hybrid} approaches. Recently, deep learning-based methods attempt to solve 3D tooth segmentation. Mask R-CNN  is applied on tooth segmentation and detection \cite{cui2019toothnet}. Modified 3D Unet  structures are  well-designed with initial dental masks or complex backbones \cite{yangdeep}. However, these methods are mostly evaluated on small or in-house datasets. It is still hard to reproduce these segmentation performances as  the dental dataset and code are not published.
\begin{figure}[h]
\centering
	\includegraphics[width=\textwidth]{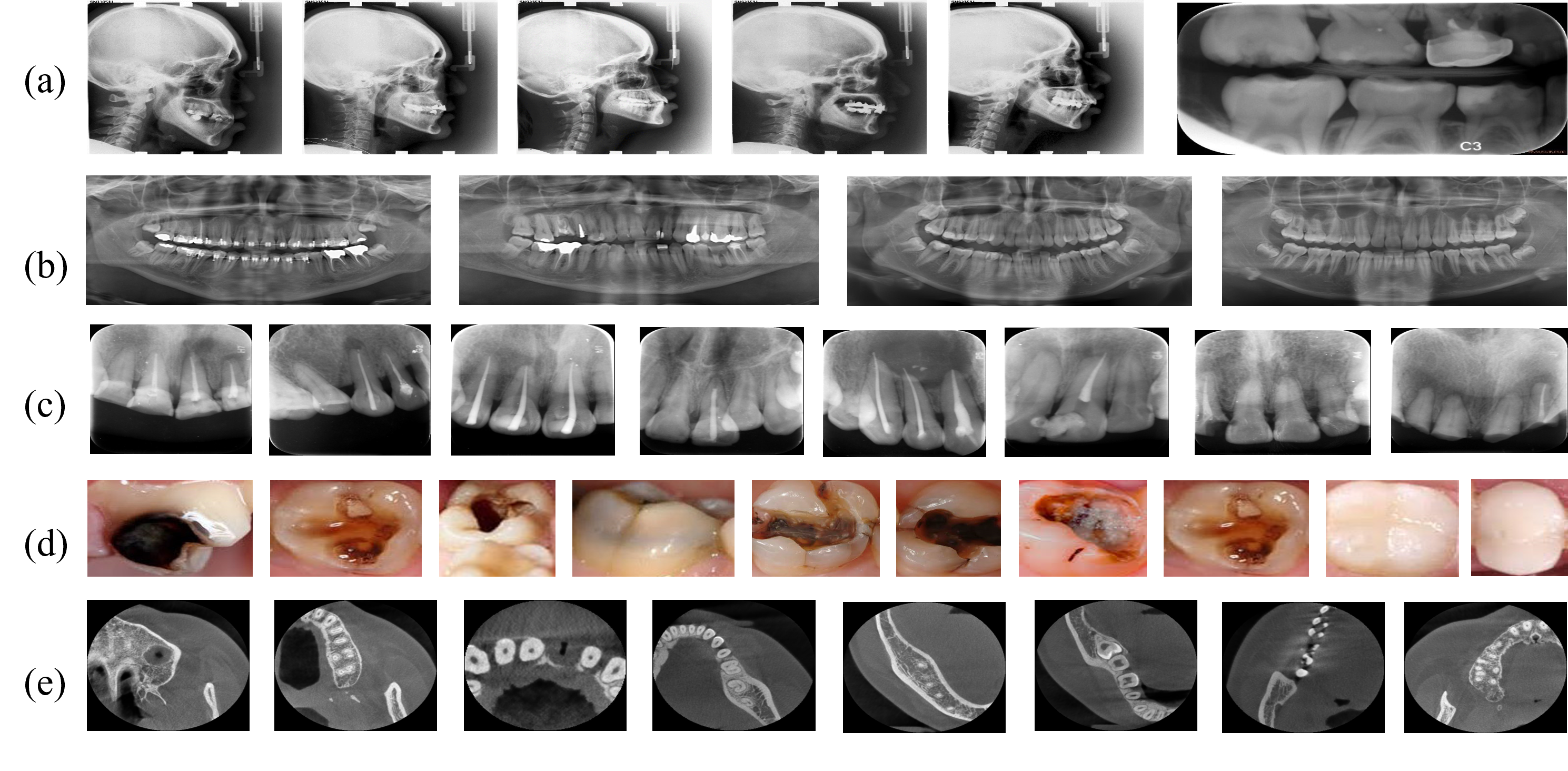}
	\caption{A few samples from different publicly available dental image datasets are illustrated.  (a) Dental X-ray Image, (b) LNDb, (c) AGMB, (d) Teeth\_dataset, and (e) CTooth+. }
	\label{fig.4}
\end{figure}

We review several  dental image datasets and summarise their contributions.  Seven types of tooth structures are marked on Dental X-ray Image dataset \cite{wang2016benchmark}.  LNDb dataset  contains polygon boundary annotations of teeth on  X-ray images \cite{silva2018automatic}. AGMB  evaluates root canal therapy on RGB images \cite{li2021agmb}. Teeth\_dataset is proposed for caries classification \cite{A:2005}. Some samples of these existing datasets are shown in Figure \ref{fig.4}. To our knowledge, no 3D dental CBCT dataset has ever been published for open-access in the medical image processing domain.    

Our work is the first comprehensive study on 3D dental data collection, annotation and evaluation.  We publish a 3D  dataset CTooth+ and release tooth segmentation performances based on fully-supervised learning, semi-supervised learning and active learning methods.  CTooth+ dataset  provides a research fundamental for following-up automatic dental segmentation studies.  


\section{CTooth+ dataset}

\subsection{Dataset Summary}
The main properties of the  existing 2D and 3D dental datasets are summarized in Table \ref{tab.1}. Compared with the published dental datasets, most of the existing datasets contain only 2D images from various tooth imaging modalities and the amount of data is relative small. Our CTooth+ fully maintains the three-dimensional characteristics of teeth, and the number of data samples exceeds 30k slices, far exceeding the existing 2D dental datasets.  The data set consists of 5504 annotated CBCT images of 22 patients and 25876 unlabeled images of 
146 patients. All patient information is coded for the purpose of protecting privacy. For each volume, we roughly spent 6 hours to annotate  tooth regions  and 
1 more hour to check and refine the annotations. In total, the CTooth+ dataset took us around 10 months to collect, annotate and review.

\begin{table}[h]
\caption{Summary of publicly available dental datasets. 
}
\centering
\setlength{\tabcolsep}{2.5mm}{
\begin{tabular}{c|cccc}
\hline
Dataset  & Year &Modality &Type   & Scan \\ \hline
Dental X-ray Image  \cite{wang2016benchmark}  & 2015 & 2D &  Bitewing       & 120 \\
LNDb \cite{silva2018automatic} & 2016 & 2D & Panomatic  X-ray     & 1,500 \\ 
Teeth\_dataset \cite{A:2005}                                                        & 2020 & 2D & Intraoral RGB image    & 77                                                \\ 
AGMB \cite{li2021agmb}&2021&2D&Root canal image&245\\ 
Our CTooth  \cite{2022arXiv220608778C}                                                        & 2022 & 3D & CBCT   & 7,368 \\ 
\textbf{Our CTooth+ }                                                           & \textbf{2022} & \textbf{3D} & \textbf{CBCT}   & \textbf{31,380} \\ \hline
\end{tabular}}
\label{tab.1}
\end{table}

 The images in CTooth+ were acquired with a OP300, manufactured by Instrumentarium Orthopantomograph$^{\circledR}$. CBCT slices were acquired in the DICOM format at the University of Electronic Science and Technology of China hospital.   All CBCT slices were scanned before dental operations, with a resolution  of 266 $\times$ 266 pixels in  the  axial  view. The in-plane  resolution  is  about  $0.25  \times  0.25  mm^{2}$ and  the slice  thickness range from 0.25 $mm$ to 0.3 $mm$. 

\begin{figure}[h]
\centering
	\includegraphics[width=\textwidth]{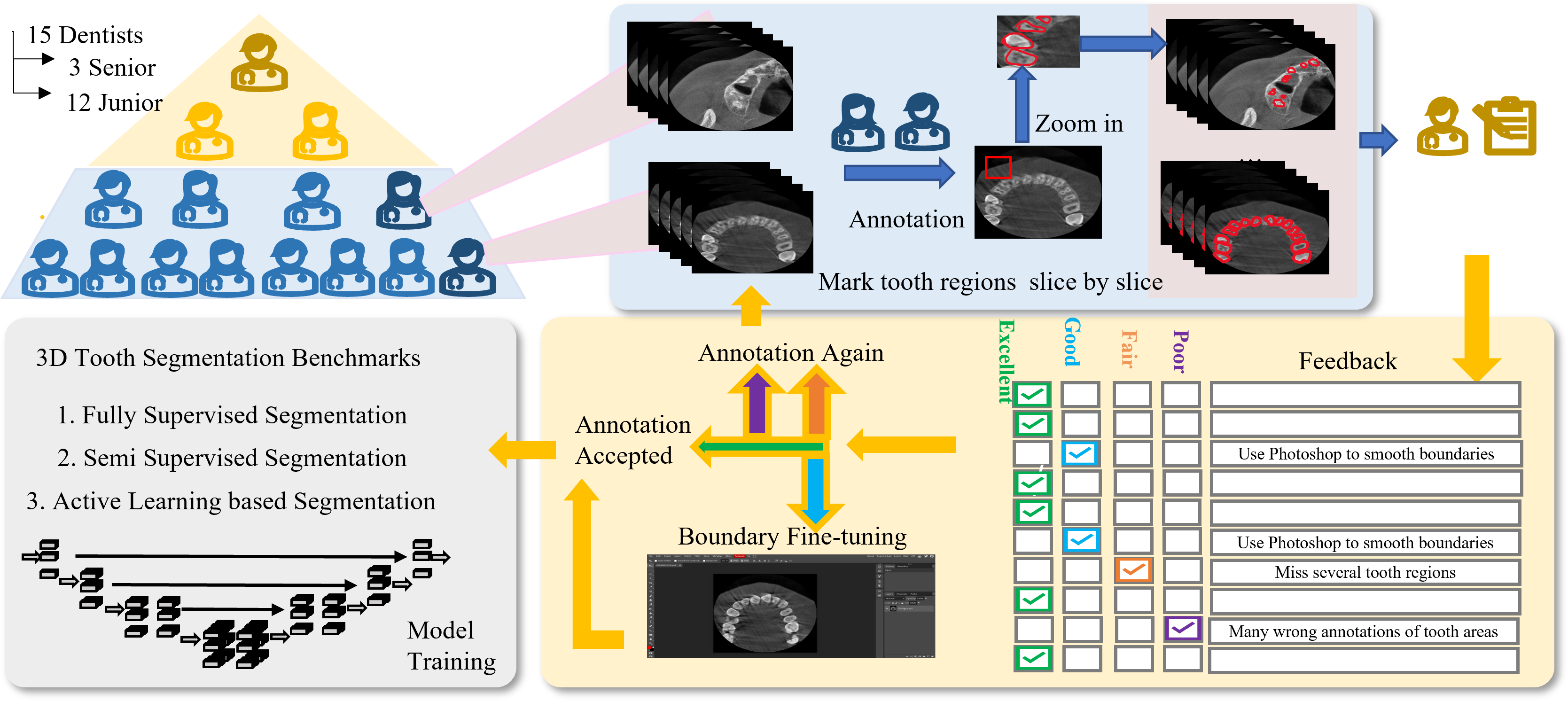}
	\caption{Dataset annotation and quality control procedure. }
	\label{fig.444}
\end{figure}

\subsection{Expert Annotation and quality assessment}

Figure \ref{fig.444} illustrates the whole procedure for CTooth+ dataset annotation and quality control procedure. Scans  were annotated by 15 dentists. Twelve junior dentists with at least two years of experience manually marked all teeth regions. They first used ITKSNAP \cite{yushkevich2006user} to delineate tooth regions slice-by-slice in the axial view. Then the annotations were modified according to the coronal view and sagittal view. 

\begin{figure}[h]
\centering
	\includegraphics[width=\textwidth]{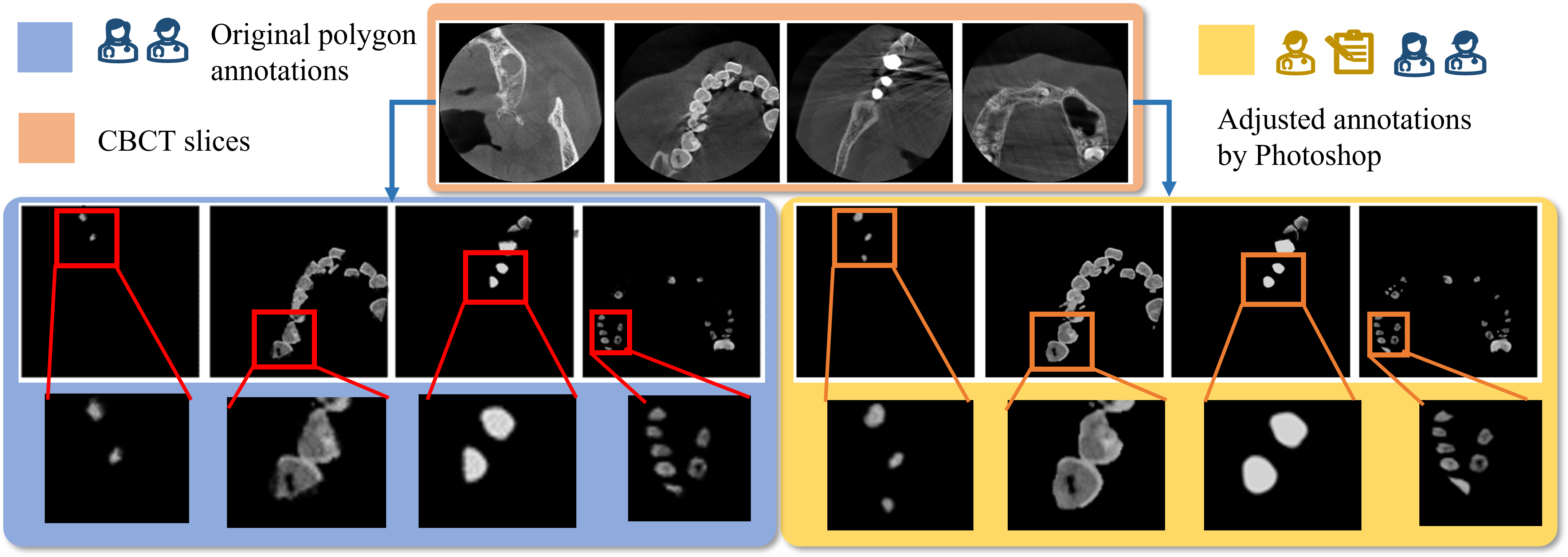}
	\caption{Annotation adjustment. }
	\label{fig.434}
\end{figure}

\begin{figure}[b]
\centering
	\includegraphics[width=\textwidth]{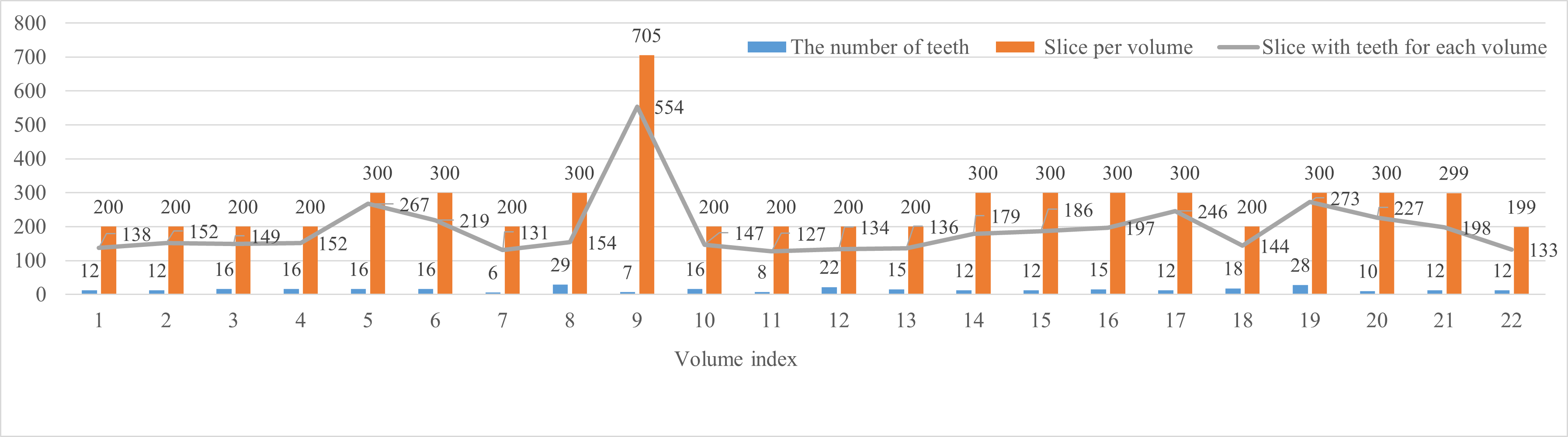}
	\caption{Annotation statistics of CTooth+.}
	\label{fig.411}
\end{figure}

Three senior experts with at least ten years of experience were invited to evaluate the tooth annotations. The senior experts assessed the annotation quality,  and marked a quality level (excellent, good, fail and poor) on each tooth annotation. ``Excellent" annotations were stored in the CTooth+ dataset directly. ``Good" annotations were fed into Phototshop software \cite{manovich2011inside} for  fine-tuning according to the experts' feedback. ``Fair" and ``Poor" annotations and their feedback were put back into the unlabelled data pool and were marked again.  In Figure \ref{fig.434}, we illustrate a set of ``Good" annotations before and after adjustment. It is clear that the tooth boundaries are more precise and smoother after necessary adjustment. 


Statistics of annotated teeth are illustrated in Figure \ref{fig.411}. All image volumes have about 12 teeth, 200 or 300 slices,  and 150 slices with teeth except for the 9$^{th}$ volume. The unlabelled images have similar statistics as the annotated images. The similar data statistics attributes ensure the stability of model training. In addition, variance in tooth shape, restorations, implants inside each volume forces the segmentation model to learn with robustness and generalizability.

\subsection{Potential research topics}

Fully-supervised learning (FSL) based segmentation efficiently exploit labelled data and solve  complex challenges, e.g. imbalanced distributions. FSL based tooth segmentation has been studied recently but no open-access dataset is published for evaluating these methods. Here, we propose the 3D dental dataset CTooth+ and reproduce eight  FSL segmentation methods based on it.

Semi-supervised learning (SSL)  requires less expert annotations for model training, relieving the time and labour burden associated to data annotation.   To our knowledge, there is no SSL-based tooth volume segmentation method published.  This work attempts to apply four state-of-the-art SSL medical segmentation methods on CTooth+ and evaluate their performances.


Compared to FSL (accurate but  expensive) and SSL (economical but affected by noise), various active learning (AL) strategies are designed to enlarge the training set by iteratively selecting informative samples. In this paper, we extend six active learning methods to their 3D version and evaluate their tooth segmentation performances on CTooth+.


\section{Experiments and results}

\subsection{Evaluation metrics and implementations on the CTooth+}
 
\subsubsection{Evaluation Metrics:}The segmentation results are evaluated  using  dice similarity coefficient (DSC), intersection-over-union (IOU), sensitivity (SEN), positive predictive value (ppv), Hausdorff  distance (HD),  average symmetric  surface  distance (ASSD),  surface overlap (SO),  and surface dice (SD) \cite{2022arXiv220608778C}. 
\subsubsection{Implementation Details:} Kaiming initialization \cite{he2015delving} is used for initializing all the weights of models.  The Adam optimizer is used with  a learning rate of 0.0004 and a step learning scheduler (with step size=50 and $\gamma=0.9$).  All networks are trained for 300 epochs using a sever with 2 Nvidia A100s and 48 GB CPU memory. All images are divided into 3D patches (size (64,128,128)) with batch size 4 to 8 according to the model complexity. We choose 20\% image volumes for evaluation and the other volumes for training the fully-supervised (with labelled images) and semi-supervised methods (with labelled and unlabelled images).  Cross entropy loss \cite{zhang2018generalized} is exploited to train all models.
\subsection{Benchmark for fully-supervised tooth volume segmentation }

We present the 3D FSL tooth segmentation performances on 8 fully-supervised segmentation methods. In Table \ref{tab.111}, Attention Unet \cite{oktay2018attention} outperforms other methods in most metrics including DSC 86.60 \%, IOU 76.45 \% and  PPV 87.79 \%, ASSD 0.27 mm, respectively. Hausdorff distance on nnUnet \cite{isensee2018nnu} is minimal at 1.29 mm, and the sensitivity metric on Voxresnet \cite{yu2017automatic} achieves the best. DenseUnet \cite{guan2019fully} has a satisfying results on the accuracy of tooth surface (SO and SD). However, we observe that 3D SkipDenseNet \cite{bui20173d} and DenseVoxelNet \cite{yu2017automatic} are both inefficient for segmenting 3D tooth volumes since their network structures are deeper than others causing network overfitting on CToooth+.
\begin{figure}[h]
\centering
	\includegraphics[width=\textwidth]{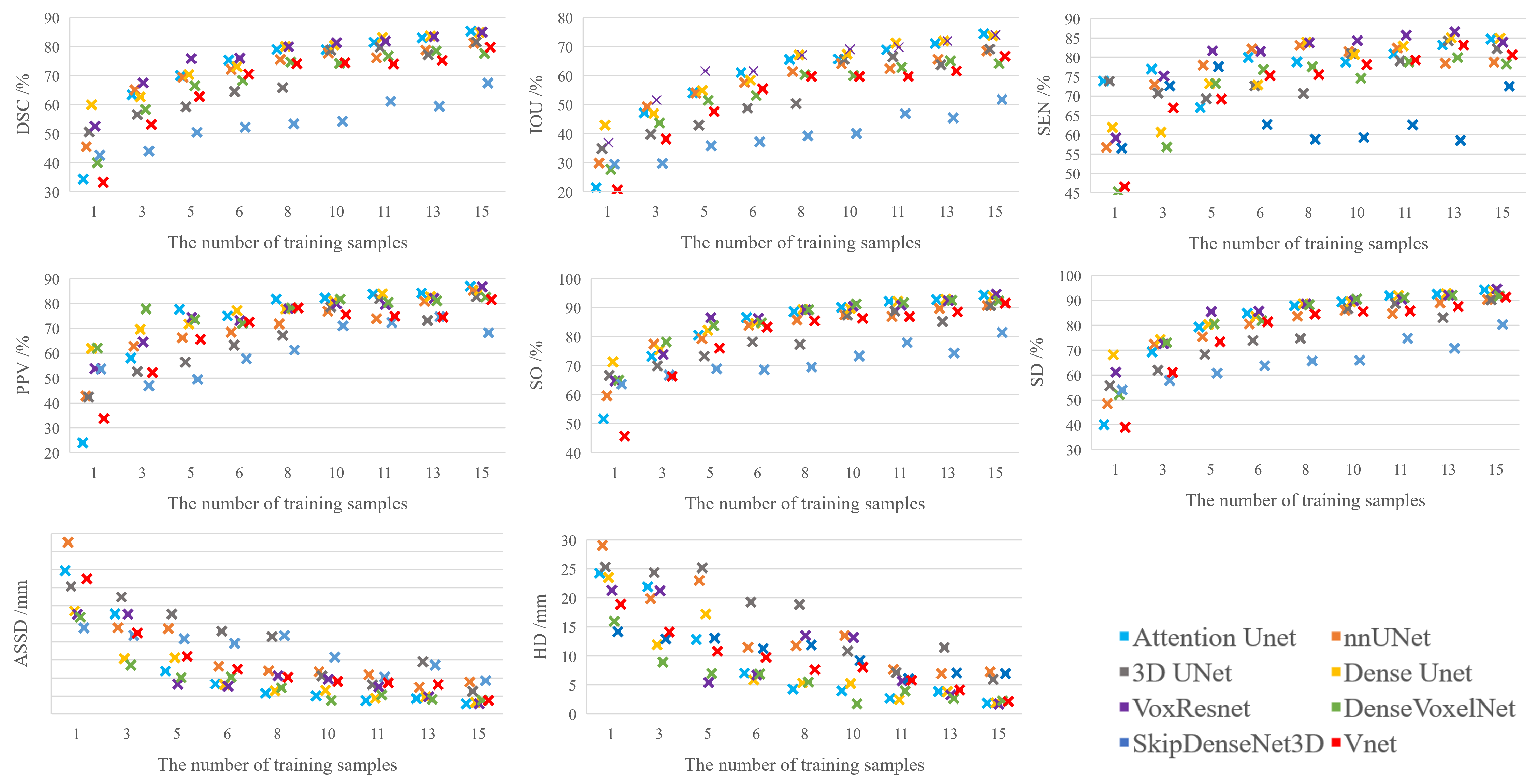}
	\caption{Evaluations on segmentation when changing the amount of training volumes.}
	\label{fig.21}
\end{figure}

\begin{table}[b]
\centering
\caption{Evaluation comparison among differnet tooth volume segmentation methods trained on 17 volumes.}
\setlength{\tabcolsep}{1.52mm}{
\begin{tabular}{c|  c c c c c c c c }
\hline
Method&DSC&IOU&SEN&PPV&HD&ASSD&SO&SD\\

\hline

3D SkipDenseNet \cite{bui20173d} &64.99&49.16&73.54&69.49&7.61&1.08&80.17&76.40\\

DenseVoxelNet \cite{yu2017automatic}&76.45&62.22&83.16&75.36&5.10&0.62&89.54&88.76\\

3D Unet \cite{cciccek20163d}&79.51 &66.40&78.21&82.78&8.02&1.01&89.22&88.76 \\

VNet \cite{milletari2016v}&81.21&68.58&80.88&83.27&1.61&0.29&93.11&92.90\\

Voxresnet \cite{yu2017automatic} &85.07&74.25&86.58&84.29&5.14&0.45&94.11&94.04 \\

nnUnet \cite{isensee2018nnu}&85.48&74.83&84.56&87.22&\textbf{1.29}&\textbf{0.27}&95.09 &95.03\\

Dense Unet \cite{guan2019fully} &86.27&76.11&\textbf{90.80}&83.23&2.08&0.39&\textbf{95.98}&\textbf{95.91} \\

Attention Unet \cite{oktay2018attention}& \textbf{86.60}&\textbf{76.45}&86.11&\textbf{87.79}&1.72&\textbf{0.27}&95.25&95.20 \\

\hline
\end{tabular}}
\label{tab.111}
\end{table}

We further perform an ablation study on the FSL tooth segmentation task. Figure \ref{fig.21} shows the quantitative segmentation performances among the FSL segmentation methods when changing the number of training sample volumes.  It is clear that all performance metrics increase  when the number of data samples increases evenly. However, noise and uneven data sampling make the increase in data volume unproportional to the performance gain. Hence, more designs are considered to increase network robustness and reduce the noise effect. 

\subsection{Benchmark for semi-supervised tooth volume segmentation}

 SSL based tooth segmentation exploits less ground truth and a large number of unlabeled images for training. In Table \ref{tab.33}, we compare the segmentation performances of four state-of-the-art SSL strategies trained by 9 labelled volumes and 8 unlabelled volumes. Experimental results show that all these SSL models achieve better performance than the FSL based Dense Unet trained on only 9 labelled volumes. CTCT \cite{luo2021semi}  outperforms others. 
 
 \begin{table}[b]
\centering
\caption{3D tooth segmentation performance comparison among 4 SSL methods. All models are trained by 9 labelled volumes. }

\setlength{\tabcolsep}{1.2mm}{


\begin{tabular}{c|c|ccccccc}
\hline
 \# Unlabeled volume    & Method            & DSC &IOU&SEN &PPV&HD&ASSD \\ \hline
 /                   & Dense Unet \cite{guan2019fully}           & 78.99&65.55&78.81&81.71&4.29 &0.57\\
\cline{1-8}
                     \multirow{4}{*}{8}  & MT \cite{tarvainen2017mean}     & 82.66&70.55&83.05&83.11&\textbf{2.76}&0.52 \\ 
                                         & CPS \cite{chen2021semi}       & 83.17&71.48&83.10&83.02&4.13&0.55 \\ 
                                        & DCT \cite{qiao2018deep}  & 83.10&71.33&83.62&83.10&4.28&0.56    \\  
                                       & CTCT \cite{luo2021semi}    & \textbf{85.32}&\textbf{74.60}&\textbf{87.55}&\textbf{84.22}&2.81&\textbf{0.43} \\ \hline
\end{tabular}}

\label{tab.33}
\end{table}

 \begin{figure}[h]
\centering
	\includegraphics[width=\textwidth]{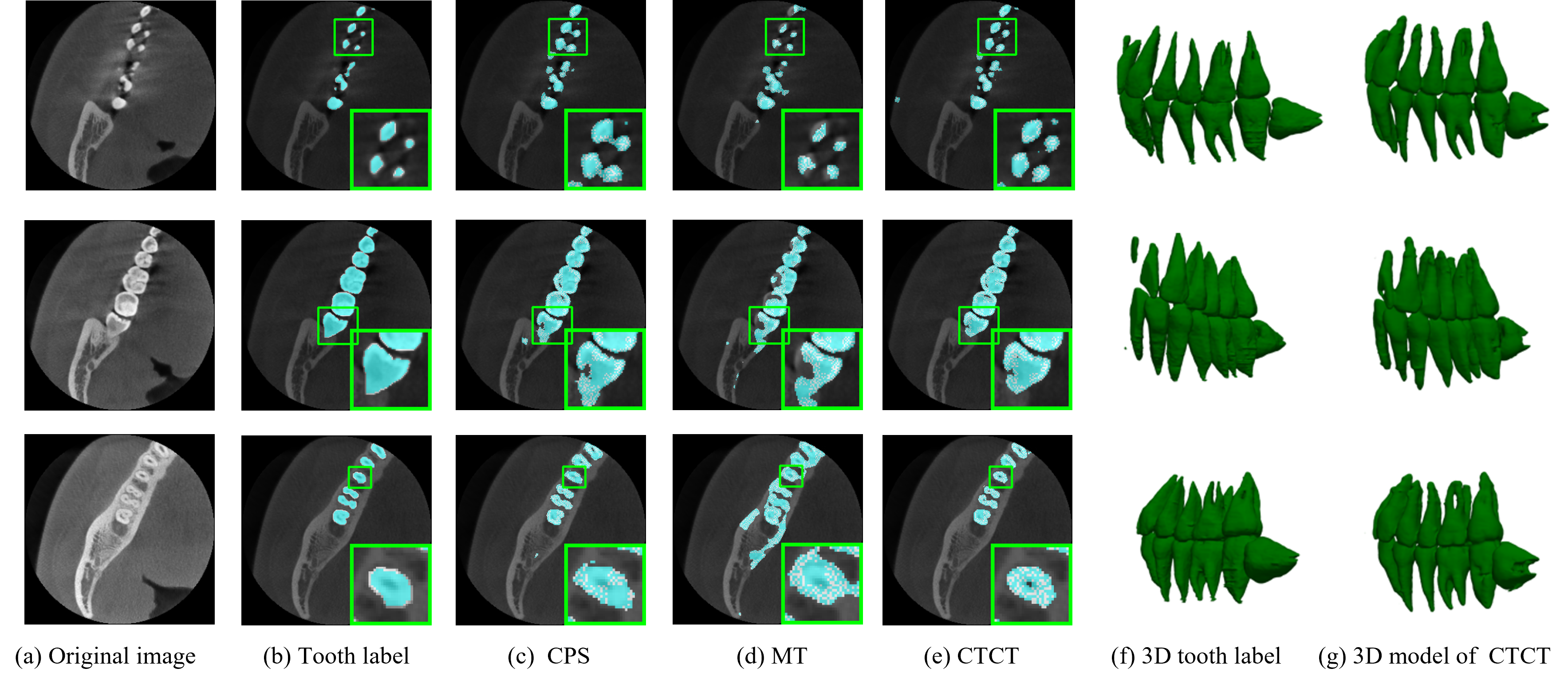}
	\caption{Qualitative SSL segmentation results.  A closer look  reveals  clear  tooth boundaries at  the right bottom corner of each slices.}
	\label{fig.400}
\end{figure}

In Figure \ref{fig.400}, we compare the segmentation details among SSL methods. CPS \cite{chen2021semi} and MT \cite{tarvainen2017mean} are not as accurate as CTCT \cite{luo2021semi} method especially in the tooth root regions. We also compare the 3D tool model based on the segmentation boundaries between ground truth and predicted results of CTCT as shown in sub-figure (f) and (g). It can be seen that the boundary details of CTCT are close to the ground truth.

\subsection{Benchmark for active learning based tooth volume segmentation}

To reduce the noise effect, we reproduce six AL based medical segmentation methods based on the Attention Unet backbone and present the performances. In Table \ref{tab.112}, CEAL \cite{hwa2004sample} achieves the comparable performances as FSL but uses 12 \% less training data. However,  ENT  \cite{hwa2004sample} and MAR  \cite{joshi2009multi} both have similar performances as the FSL when they are trained on 72 
patches. These experiments present that active learning-based tooth volume segmentation is effective but still needs more designs to explore tooth information representation.

\begin{table}[h]
\centering
\caption{Evaluation comparison among differnet tooth volume segmentation methods.}
\setlength{\tabcolsep}{0.8mm}{

\begin{tabular}{c|c|cccccccc}
\hline
\# 3D  Patches     & AL strategy      & DSC   & IOU   & SEN   & PPV   & HD   & ASSD & SO    & SD    \\ \hline
56                        & \textbackslash{} & 81.44 & 68.86&80.88&83.73&2.71&0.37&92.12&91.85
\\ 
72&\textbackslash{} &85.28&	74.41&	84.69&	86.90&	1.88&	0.28&	94.28&	94.20\\ 
82&\textbackslash{}&\textbf{86.60}&\textbf{76.45}&86.11&\textbf{87.79}&1.72&0.27&95.25&95.20 \\\hline
\multirow{3}{*}{72} 

                          & ENT  \cite{hwa2004sample}     &      83.92&	72.49&	82.44&	86.36&	1.42&	0.27&	94.21&	94.14 \\

                           & MAR  \cite{joshi2009multi}            & 84.88&	73.86&	83.30&87.30	&	1.63&	0.29&	94.08&	94.03
 \\
 & CEAL \cite{wang2016cost} &86.58&	76.43&	\textbf{87.85}&	86.01&\textbf{1.05}&	\textbf{0.21}&	\textbf{95.92}&	\textbf{95.89}
       \\
       \hline
\end{tabular}}

\label{tab.112}
\end{table}

\section*{Acknowledgement}
The work was supported by the  the Natural Science Foundation of China under Grant No. 62002316.



\section{Conclusion}
This work is the first to collect and publish a 3D dental dataset CTooth+ with annotated 3D structures of teeth according to quality assessment from experts, and evaluate the tooth volume segmentation on FSL, SSL and AL methods systematically as benchmarks. In future, we will release more data from multiple dental organisations and release more annotations on the tooth structures.
%
%
%

\bibliographystyle{splncs04}
\bibliography{reference}

%




\end{document}